# Strong $T_c$ dependence for strained, epitaxial $Ba(Fe_{1-x}Co_x)_2As_2$ thin films

K. Iida[1], J. Hänisch[1], R. Hühne[1], F. Kurth[1], M. Kidszun[1], S. Haindl[1], J. Werner[2], L. Schultz[1] and B. Holzapfel[1]

[1] IFW Dresden, Institute for Metallic Materials (IMW), *P.O. Box 27 01 16, 01171 Dresden, Germany*
[2] IFW Dresden, Institute for Solid State Research (IFF), *P.O. Box 27 01 16, 01171 Dresden, Germany*

$Ba(Fe_{1-x}Co_x)_2As_2$ superconducting thin films have been grown on $SrTiO_3$, $(La,Sr)(Al,Ta)O_3$, $LaAlO_3$ and $YAlO_3$ (YAO) single crystal substrates by pulsed laser deposition. All the films, except on YAO, have been grown epitaxially without buffer layers. The films deposited on YAO contained 45° in-plane rotated grains and showed a broad superconducting transition. The onset $T_c$ of the films is observed to increase from 16.2 K to 24.5 K with increasing $c/a$, mainly due to a slight distortion of the $AsFe_4$ tetrahedron. From this correlation, we expect that higher superconducting transition temperatures than 24.5 K in a strained epitaxial film may be possible.

The discovery of the superconducting fluorine-doped, so-called "1111" compound with the chemical formula of $LaFeAsO_{1-x}F_x$ has triggered worldwide research efforts in both theoretical and experimental aspects for the purpose of understanding the physics and discovering higher-$T_c$ materials.[1] As a result, a record $T_c$ of 43 K has been achieved immediately in the "1111" compound by replacing the lanthanum ion with samarium,[2] followed by the optimised corresponding compound with a record $T_c$ of 55 K.[3] Soon after the discovery of the "1111" compound, Rotter *et al.* have pointed out that $BaFe_2As_2$ (hereafter "122"), which belongs to the $ThCr_2Si_2$ structure type, possesses a similar FeAs layer like the "1111" compound, leading to the discovery of potassium containing $BaFe_2As_2$ with a high $T_c$ of 38 K.[4] On the analogy of the FeAs layer in the "1111" compounds, both LiFeAs ("111") and FeSe ("11") iron pnictides have been also discovered to be superconducting.[5-6] Here it is noted that both "111" and "11" compounds show superconductivity without elemental substitution, since the final products of these compounds have always been Se and/or As deficient to date (i.e. off stoichemetry) leading to a natural hole doping. In both "1111" and "122" mother compounds, a spin density wave anomaly together with a structural phase transition has been observed. Both compound classes do not show superconductivity unless adding holes and/or electrons or applying hydrostatic pressure.[7-8] Here, most of the research has been carried out on either bulk single crystals or polycrystalline materials.

Epitaxial thin films are suitable for investigating the uniaxial pressure dependence of superconducting properties since either contraction or expansion can be induced into the film by the lattice misfit within a critical thickness of stress relaxation. However, there have been difficulties in fabricating superconducting iron pnictide films, especially "1111" compounds.[9-10] Nevertheless, superconducting F-doped LaFeAsO thin films have been successfully deposited by our group[11] (quite recently epitaxial films[12]) and their detailed characterization can be found in Ref. 13. Among the family of iron pnictides, epitaxial superconducting "122" and "11" films have been readily fabricated by a simple pulsed laser deposition technique,[14-17] albeit the latter and the Sr-based former materials have been reported to be unstable in ambient atmosphere. In this letter, we report on the deposition of $Ba(Fe_{1-x}Co_x)_2As_2$ films on various substrates and the effect of the lattice misfit on their superconducting properties.

The $Ba(Fe_{0.9}Co_{0.1})_2As_2$ target for pulsed laser deposition was prepared by a conventional solid state reaction process. A stoichiometric mixture of $Ba_5As_3$, $Fe_2As$, FeAs, and Co was pressed into a pellet of 10 mm in diameter and 5 mm in height and placed in a glassy carbon crucible, which did not react with the sample during the process. The whole arrangement was put into a quartz tube and sealed under Ar atmosphere. The sample was heated to 900 °C at a rate of 300 °C/h, held at this temperature for 16 h, and then immediately cooled to room temperature. The phase purity of the target was determined by XRD using a standard Bragg-Brentano geometry with Co-$K_\alpha$ radiation. All observed peaks were indexed with the 122 phase except a small amount of FeAs impurity phase. The refined lattice parameters, using the program RIETAN-FP, were $a = 3.9582(9)$ (Å), $c = 12.988(2)$ (Å) and $z = 0.3578(8)$ (atomic position of As), respectively.[18] As compared with the reference data,[19] our target material showed a slight decrease in the length of $a$–axis and an increase of the $c$-axis. Based on the refined lattice parameters, the As-Fe-As bond angle $\alpha$ in the $FeAs_4$ tetrahedron was evaluated to 109.5(2)°, which is the almost identical value of a regular tetrahedron ($\alpha$=109.47°).

$Ba(Fe_{1-x}Co_x)_2As_2$ films with an approximate thickness of 40 nm were deposited on various (001)-orientated cubic or pseudo-cubic substrates such as $SrTiO_3$ (STO), $(La,Sr)(Al,Ta)O_3$ (LSAT), $LaAlO_3$(LAO) and $YAlO_3$ (YAO) by a pulsed laser deposition technique, where the $Ba(Fe_{0.9}Co_{0.1})_2As_2$ target was ablated with 248 nm KrF radiation at a frequency of 20 Hz under UHV conditions (base pressure $10^{-9}$ mbar). Prior to every deposition, the substrate was heated to 800 °C, held at this temperature for 20 min, subsequently cooled to 650 °C (for LSAT, 700 °C) followed by another 20 min holding time. For the STO substrate, a high-temperature treatment of 1000 °C and a longer holding time of 1 h were employed in order to remove oxygen from the lattice since outgasing oxygen may react with As leading to an As deficient $Ba(Fe_{1-x}Co_x)_2As_2$ film. The energy density of the laser on the target was around 3–5 J/cm$^2$, and the distance between target and substrate was 50 mm. The films grow with a very smooth surface with an RMS roughness of 11.9 nm, as measured by atomic force microscopy. EDS chemical analyses revealed no appreciable change in each deposited film, which infers that the influence of chemical fluctuation on the following results is negligible.

XRD patterns in logarithmic scale (fig. 1) clearly show that all the films have been grown with high phase purity and *c*-axis texture, i.e. with [001] perpendicular to the substrate surface. The respective lattice constants *c* of the films deposited on STO, LSAT, LAO and YAO were evaluated by using the Nelson-Riely function to 13.111(2) Å, 13.017(3) Å, 13.002(1) Å, and 12.907(4) Å. Calculated lattice parameters *c* of the films on STO and LSAT are apparently larger than the target $Ba(Fe_{0.9}Co_{0.1})_2As_2$ value of 12.988(2) Å, indicating a lattice tensile strain along [001]. The film on LAO also showed a slight increase in length of *c*-axis. In stark contrast, the lattice constant *c* of the films deposited on YAO is smaller than the bulk value.

The (103) pole figure measurement of the $Ba(Fe_{1-x}Co_x)_2As_2$ film on STO showed a clear four-fold symmetry and a full width at half-maximum (FWHM) of 0.8°, indicating that the films have been grown epitaxially with the relation (001)[100] || (001)[100]. The films on LSAT and LAO were also grown epitaxially, however, the FWHM value $\Delta\phi$ was slightly increased to 1.4° and 1.3°. On the other hand, the films on YAO contained 45 ° rotated-grains, as shown in fig 2 (b).

The relationship between the in-plane lattice parameters, *a*, of each film, which was evaluated by using the data of high resolution reciprocal space maps, and the out-of-plane lattice parameters, *c*, is shown in fig. 3. It is apparent that the lattice constant *c* for the films is observed to decrease linearly with increasing lattice constant *a*. Both lattice parameters, *a* and *c*, of the film on LAO are almost identical to the bulk value, albeit a relatively large misfit of -4.25 %. On the other hand, the lattice constants of the film on STO with a small misfit of -1.34 % differ from the bulk value. These results can be explained in terms of the coherent growth. According to this theory, the film on STO possesses the thickest critical layer (hence strained layer) among the films, however, the strain can be relaxed by increasing the thickness of the film. In order to verify the above discussion, the lattice constant of a thicker film of 90 nm on LSAT has been examined. As a result, the lattice constant *c* of the resultant film was evaluated to 12.954(2) Å, which is close to the bulk value, indicating that the film is relaxed. Furthermore, the onset $T_c$ was decreased from 23 K to 21.5 K. The film on YAO with the largest misfit of -6.27 % was also strained, however, it was partially relaxed due to the grain boundaries.

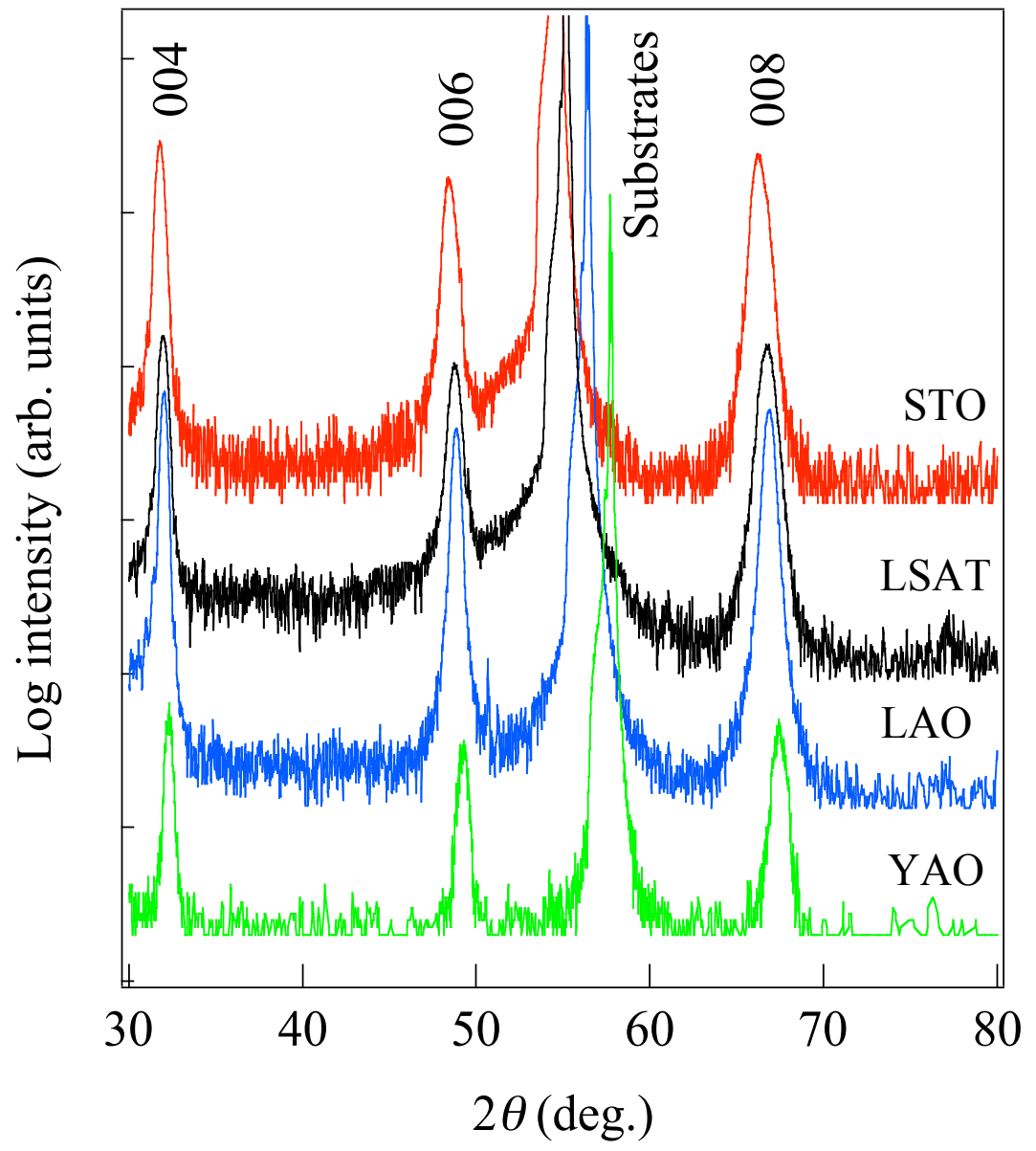

FIG. 1 (Color online) Only 00*l* peaks were observed in all the films on STO, LSAT, LAO and YAO. The 200 substrate peak of each film was observed in the range of 50° to 60°. The data were recorded in a standard Bragg-Brentano geometry with Co-$K_\alpha$ radiation.

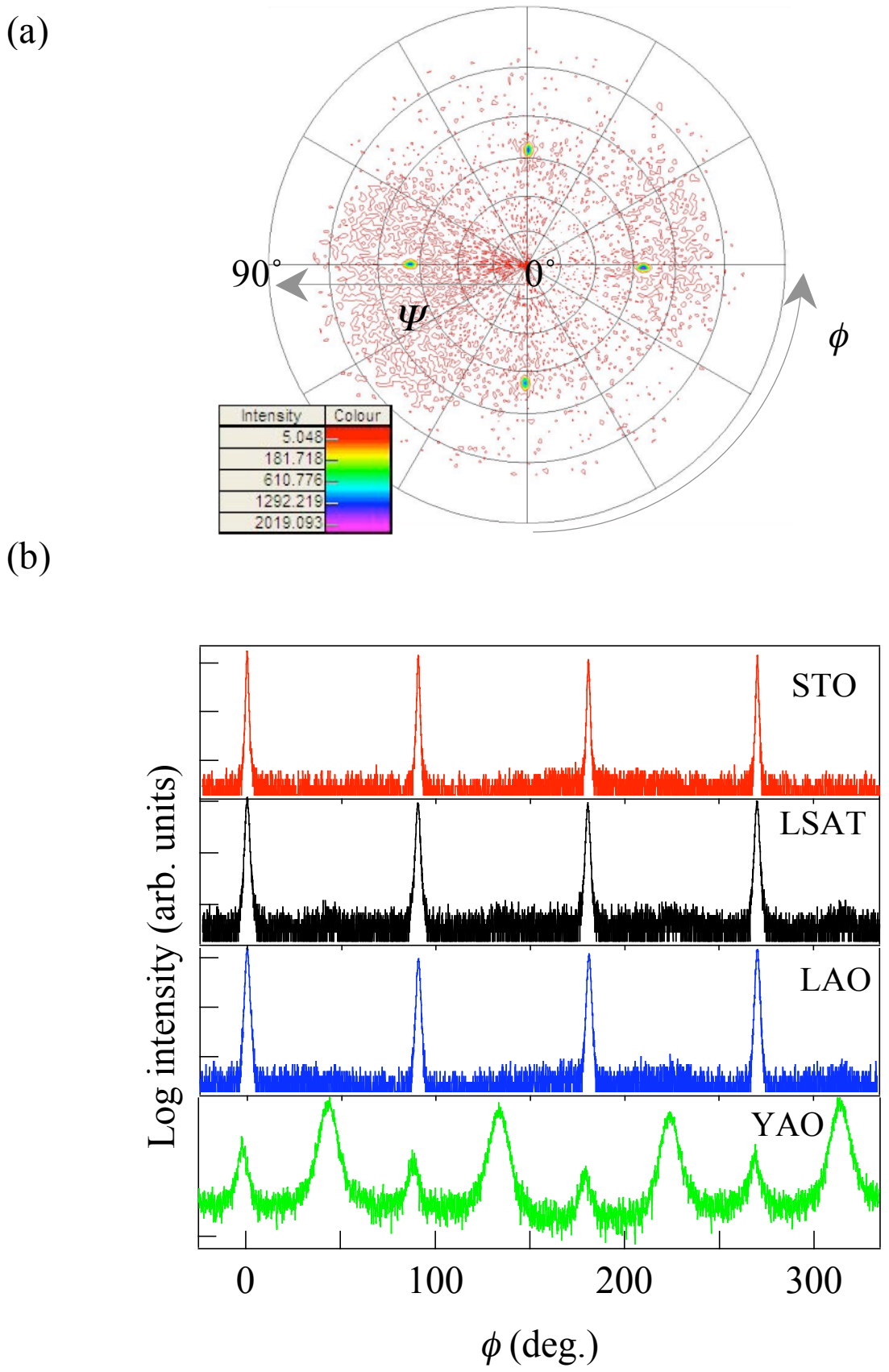

FIG 2 (Color online) (a) Representative (103) pole figure of a Co-containing $BaFe_2As_2$ thin film on STO showed a clear four-hold symmetry. The film on YAO substrate contains 45° rotated grains, as shown in figure 2(b). The data were recorded in a texture goniometer system operating with Cu-$K_\alpha$ radiation.

A wide superconducting transition was observed for the films on YAO compared with the other films as shown in fig. 4(a), which is owing to weak-link behaviour. The resistivity of the film on STO exhibited a quite low value since the STO substrate becomes conducting under the applied deposition conditions. The onset $T_c$ of the films was observed to increase with increasing *c*/*a*, as shown in fig. 4(b). It is also notable from fig. 4(b) that superconductivity at higher temperatures than 24.5 K for epitaxial $Ba(Fe_{1-x}Co_x)_2As_2$ films might be possible once the value of *c*/*a* is further increased (hence strained). From the above results, the lattice distortion clearly affects the superconducting properties. This is in contrast to the fact that $BaFe_2As_2$ bulk single crystals with optimal Co content show no significant influence on $T_c$ under applied hydrostatic pressure.[20] It was reported that $T_c$ in the iron pnictide family significantly varies with the As-Fe-As bond angle, $\alpha$, in the $FeAs_4$ tetrahedron.[21] The relationship between $\alpha$ and $T_c$ shows a quadratic dependence, and the highest $T_c$ value of iron pnictides is observed for undistorted $FeAs_4$ tetrahedra with an $\alpha$ value of around 109.47°.[21] The As-Fe-As bond angles in the present films on LAO and YAO substrates differ from 109.47° due to the lattice deformation. Accordingly, these films exhibit a lower $T_c$ compared with films on STO and LSAT.

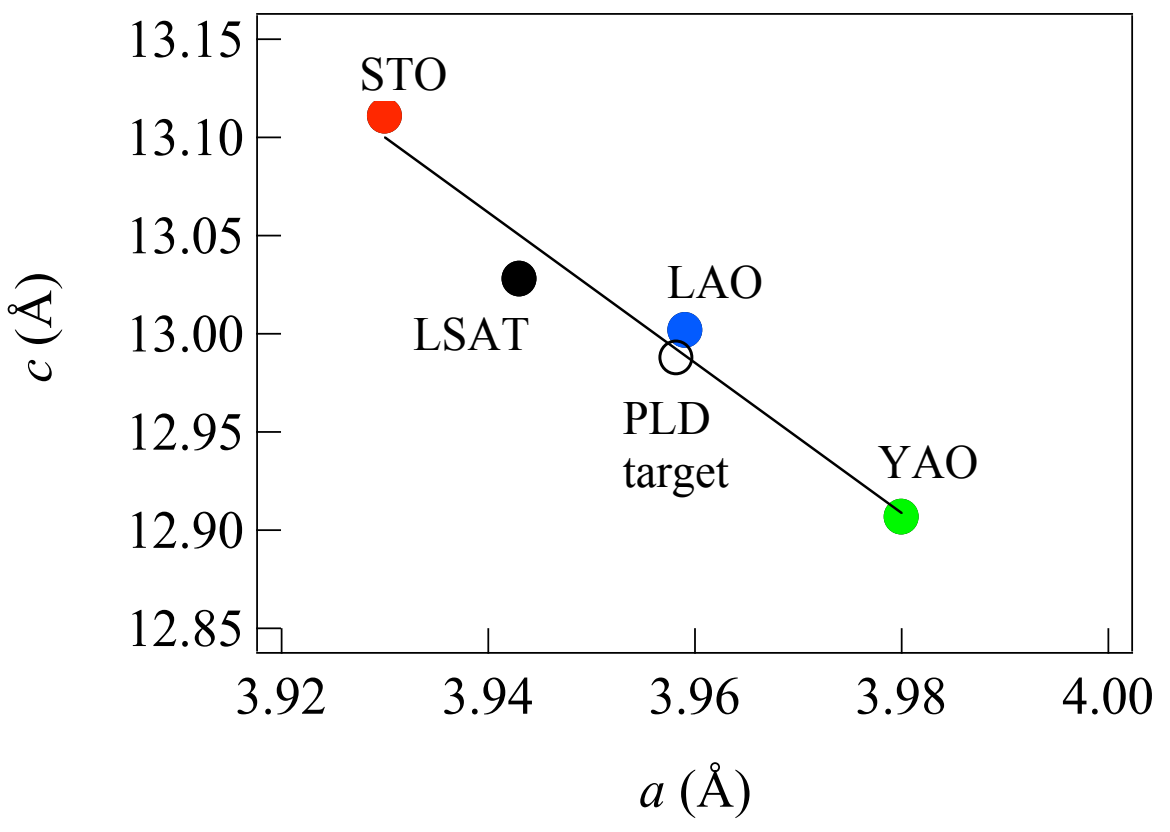

FIG. 3 (Colour online) The lattice parameter $c$ decreases linearly with increasing $a$ for the grown films on STO, LSAT, LAO, and YAO. The lattice parameters, $a$ and $c$, of the film on LAO are almost identical value of the bulk value.

In summary, we have systematically investigated the effect of lattice misfit on the superconducting properties of $Ba(Fe_{1-x}Co_x)_2As_2$ thin films by employing various single crystal substrates. All the films except on YAO have been grown epitaxially. The films on YAO, however, contained 45˚ rotated grains. A strong correlation between the lattice deformation and $T_c$ has been found, which can be explained in terms of the structural distortions of the $FeAs_4$ tetrahedra within the crystal structure. The possibility of superconductivity at higher temperatures than 24.5 K for epitaxial $Ba(Fe_{1-x}Co_x)_2As_2$ thin films has been proposed. The key to tune $T_c$ in epitaxial thin films is the precise control of the lattice misfit.

**Acknowledgements**

The authors would like to thank M. Deutschmann, C. Nacke, K. Tscharntke and U. Besold for their technical support.

## References

[1] Y. Kamihara, T. Watanabe, M. Hirano, H. Hosono, J. Am. Chem. Soc. **130**, 3296 (2008).
[2] X.H. Chen, T. Wu, G. Wu, R.H. Liu, H. Chen, D.F. Fang, Nature **453**, 76 (2008).
[3] Z.A. Ren, W. Lu, J. Yang, W. Yi, X.L. Shen, Z.C. Li, G.C. Che, X.L. Dong, L.L. Sun, F. Zhou and Z.X. Zhao, Chin. Phys. Lett. **25**, 2215 (2008).
[4] M. Rotter, M. Tegel and D. Johrendt, Phys. Rev. Lett. **101**, 107006 (2008).
[5] J.H. Tapp, Z.J. Tang, B. Lv, K. Sasmal, B. Lorenz, P.C.W. Chu, A.M. Guloy. Phys. Rev. B **78**, 060505 (2008).
[6] F.C. Hsu, J.Y. Luo, K.W. Yeh, T.K. Chen, T.W. Huang, P.M. Wu, Y.C. Lee, Y.L. Huang, Y.Y. Chu, D.C. Yan, M.K. Wu, PNAS **105**, 14262 (2008).
[7] H. Takahashi, H. Okada, K. Igawa, K. Arii, Y. Kamihara, S. Tatsuishi, M. Hirano, H. Hosono, K. Matsubayashi and Y. Uwatoko, J. Phys. Soc. Jpn. **77**, Suppl. C, pp. 78-83 (2008).
[8] M.S. Torikachvili, S.L. Bud'ko, N. Ni and P.C. Canfield, Phy. Rev. Lett. **101**, 057006 (2008).
[9] H. Hiramatsu, T. Katase, T. Kamiya, M. Hirano and H. Hosono, Appl. Phys. Lett. **93**, 162504 (2008).
[10] T. Kawaguchi, H. Uemura, T. Ohno, R. Watanabe, M. Tabuchi, T. Ujihara, K. Takenaka, Y. Takeda, and H. Ikuta, Appl. Phys. Express **2**, 093002 (2009).

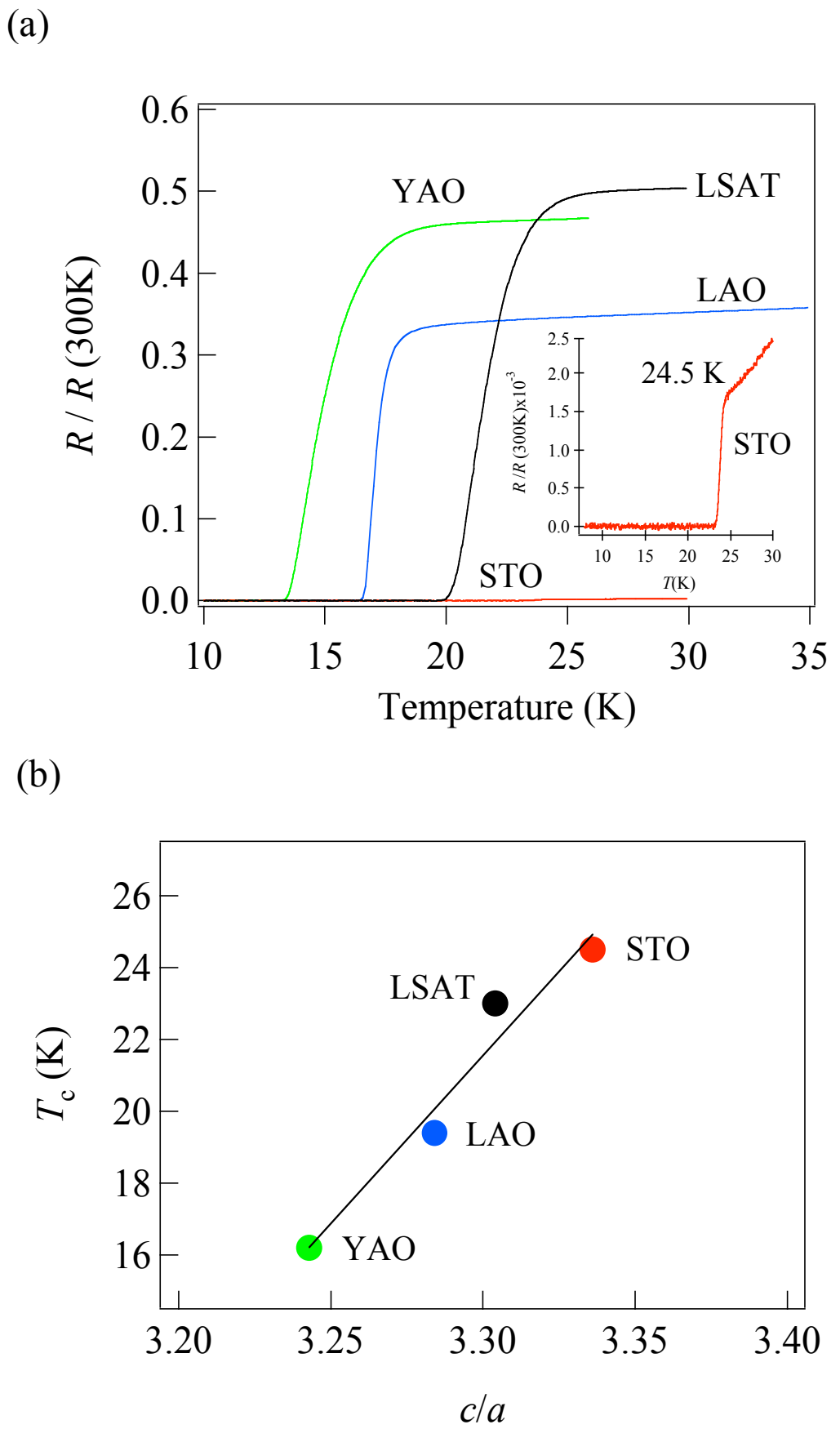

FIG. 4 (Colour online) (a) Temperature dependence of resistivity for the $Ba(Fe_{1-x}Co_x)_2As_2$ films on various substrate materials. The measurements were carried out by a standard four-probe method, in which a dc current of 1mA was employed. A wide transition width of over 3 K was observed for the films on YAO, whereas the film on STO substrate shows an onset $T_c$ of 24.5 K with a sharp transition width of less than 2 K, as shown the inset. (b) Relationship between $T_c$ and the $c/a$ shows a linear dependence. $T_c$ is significantly affected by the lattice distortion.

[11] E. Backen, S. Haindl, T. Niemeier, R Hühne, T. Freudenberg, J. Werner, G. Behr, L. Schultz and B. Holzapfel, Supercond. Sci. Technol. **21**, 122001 (2008).
[12] M. Kidzun, E. Reich, S. Haindl, J. Hänisch, B. Holzapfel and L. Schultz, arXiv:0909.3788
[13] S. Haindl, M. Kidszun, A. Kauffmann, K. Nenkov, N. Kozlova, J. Freudenberger, T. Thersleff, J. Werner, E. Reich, L. Schultz and B. Holzapfel, arXiv:0907.2271
[14] H. Hiramatsu, T. Katase, T. Kamiya, M. Hirano and H. Hosono, Appl. Phys. Express **1**, 101702 (2008).
[15] M.K. Wu, F.C. Hsu, K.W. Yeh, T.W. Huang, J.Y. Luo, M.J. Wang, H.H. Chang, T.K. Chen, S.M Rao, B.H. Mok, C.L. Chen, Y.I. Huang, C.T. Ke, P.M. Wu, A.M. Chang, C.T. Wu and T.P. Perng, Physica C **469**, 340 (2009).
[16] T. Katase, H. Hiramatsu, H. Yanagi, T. Kamiya, M. Hirano and H. Hosono, arXiv:0907.0666
[17] S. Lee, J. Jiang, J.D. Weiss, C.M. Folkman, C.W. Bark, C. Tarantini, A. Xu, D. Abraimov, A. Polyanskii, C.T. Nelson, Y. Zhang, S.H. Baek, H.W. Jang, A. Yamamoto, F. Kametani, X.Q. Pan, E.E. Hellstrom, A. Gurevich, C.B. Eom and D.C. Larbalestier, arXiv:0907.3741
[18] F. Izumi and K. Momma, Solid State Phenom. **130**, 15 (2007).
[19] A.S. Sefat, R. Jin, M.A. McGuire, B.C. Sales, D.J. Singh and D. Mandrus, Phys. Rev. Lett. **101**, 117004 (2008).
[20] K. Ahilan, J. Balasubramaniam, F.L. Ning, T. Imai, A.S. Sefat, R. Jin, M.A. McGuire, B.C. Sales and D. Mandrus, J. Phys.: Condens. Matter **20**, 472201 (2008).
[21] C.H. Lee, A. Iyo, H. Eisaki, H. Kito, M. Teresa Fernandez-Diaz, T. Ito, K. Kihou, H. Matsuhata, M. Braden and K. Yamada, J. Phys. Soc. Jpn. **77**, 083704 (2008).